# 3D simulations of TRAPPIST-1e with varying $CO_2$, $CH_4$ and haze profiles

Mei Ting Mak,[1]★ Denis Sergeev,[1] Nathan Mayne,[1] Nahum Banks,[1] Jake Eager-Nash,[1] James Manners,[2,1] Giada Arney,[3] Éric Hébrard,[1] Krisztian Kohary[1]

[1]*Department of Physics and Astronomy, University of Exeter, Exeter, UK*
[2]*Met Office, Exeter, UK*
[3]*NASA Goddard Space Flight Center, Greenbelt, MD, USA*

**ABSTRACT**
Using a 3D General Circulation Model, the Unified Model, we present results from simulations of a tidally-locked TRAPPIST-1e with varying carbon dioxide $CO_2$ and methane $CH_4$ gas concentrations, and their corresponding prescribed spherical haze profiles. Our results show that the presence of $CO_2$ leads to a warmer atmosphere globally due to its greenhouse effect, with the increase of surface temperature on the dayside surface reaching up to ∼14.1 K, and on the nightside up to ∼21.2 K. Increasing presence of $CH_4$ first elevates the surface temperature on the dayside, followed by a decrease due to the balance of tropospheric warming and stratospheric cooling. A thin layer of haze, formed when the partial pressures of $CH_4$ to $CO_2$ ($pCH_4/pCO_2$) = 0.1, leads to a dayside warming of ∼4.9 K due to a change in the water vapour $H_2O$ distribution. The presence of a haze layer that formed beyond the ratio of 0.1 leads to dayside cooling. The haze reaches an optical threshold thickness when $pCH_4/pCO_2$ ∼0.4 beyond which the dayside mean surface temperature does not vary much. The planet is more favourable to maintaining liquid water on the surface (mean surface temperature above 273.15 K) when $pCO_2$ is high, $pCH_4$ is low and the haze layer is thin. The effect of $CO_2$, $CH_4$ and haze on the dayside is similar to that for a rapidly-rotating planet. On the contrary, their effect on the nightside depends on the wind structure and the wind speed in the simulation.

**Key words:** planets and satellites: atmospheres - planets and satellites: terrestrial planets - radiative transfer - planets and satellites: dynamical evolution and stability

## 1 INTRODUCTION

So far more than 5,500 exoplanets have been detected[1]. To determine whether these exoplanets may support habitable conditions, for example, by sustaining liquid water on their surface, it is important to first understand that the modern Earth is not the only example of a life-supporting climate. During the Archean period, which refers to the Earth 2.5–4 billion years ago (Knoll & Nowak 2017), and represents around 40% of the entire Earth's history, Earth was host to the relatively simple single celled lifeforms. Whereas, the modern Earth atmosphere with a molecular oxygen $O_2$ concentration of around 21% comprises a much smaller fraction of Earth's lifetime and complex photosynthetic and oxygen-breathing life required several comparatively "unlikely" evolutionary steps (Lenton & Watson 2011). Therefore, our search for life on exoplanets might be best focused on those with atmospheric compositions similar to the Archean Earth (Eager-Nash et al. 2024).

The Archean atmosphere is thought to have been weakly reducing due to its low concentration of $O_2$, with the dominant gas species being nitrogen $N_2$, carbon dioxide $CO_2$ and $CH_4$ (Catling & Zahnle 2020). There are many studies exploring the effect of varying $CH_4/CO_2$ concentrations on the dynamics and the habitability of the Archean climate (Haqq-Misra et al. 2008; Charnay et al. 2013; Byrne & Goldblatt 2015; Goldblatt et al. 2021; Eager-Nash et al. 2023). Recently Eager-Nash et al. (2023) used the 3D General Circulation Model (GCM), the Unified Model (UM) to study the climatic impact of varying $CH_4/CO_2$ gas concentrations on the Archean Earth and found that maximum surface warming occurred when the ratio of the two partial pressures for $CH_4$ and $CO_2$ ($pCH_4/pCO_2$) was ∼0.1 (see discussion in Eager-Nash et al. 2023). However, Trainer et al. (2004, 2006) have found from their laboratory experiments that hydrocarbon hazes started forming for $CH_4/CO_2$∼0.1. Arney et al. (2016) studied the impact of hazes by coupling their 1D photochemical model to their 1D radiative and climate model, collectively called Atmos. They found the haze that formed in the photochemical model for $CH_4/CO_2$ ∼0.1, following the experiments of Trainer et al. (2004, 2006), did not have a significant radiative impact on the climate. The surface temperature remained then very similar to the haze-free case in their simulations. Yet, for $CH_4/CO_2$∼0.2, the photochemical model formed a thick haze layer that acted as a radiation shield and led to a significant cooling of the surface. Other 1D studies have also shown similar results, concluding that haze either has no radiative effect or a strong anti-greenhouse cooling effect depending on the thickness of the haze layer (Pavlov et al. 2001; Haqq-Misra et al. 2008; Zerkle et al. 2012). Mak et al. (2023) extended the work of Arney et al. (2016) and Eager-Nash et al. (2023) by using an updated version of the chemical scheme from Arney et al. (2016) and prescribing a haze layer into the same 3D GCM setup as that described in Eager-Nash et al. (2023). Mak et al. (2023) found that the thin haze layer formed when pCH$_4$/pCO$_2$ was 0.1 led to global warming of ~10.6 K. This was due to the haze absorption of shortwave radiation and re-emission of longwave radiation resulting in an increase in the atmospheric water vapour and a reduction in cloud coverage. However, they found that the thick haze layer formed when pCH$_4$/pCO$_2$ > 0.1 led to a global cooling of up to ~65 K. This was due to the strong extinction effect of haze, which prevented most of the shortwave radiation incident at the top of the atmosphere from reaching the planetary surface. On exoplanets, the irradiation regime is different to that on Archean Earth, so it is of interest to study the potential habitability of exoplanets with Archean-like atmospheric compositions, across varying CH$_4$/CO$_2$ gas ratios to include the impact of haze.

★ E-mail: M.Mak@exeter.ac.uk
[1] http://exoplanet.eu

To choose the appropriate candidate for such study, we first note that most stars in our galaxy are M-dwarfs (Rodono 1986). Morton & Swift (2014) estimated that for every ten M-dwarfs, there are potentially eight Earth-sized exoplanets orbiting in the habitable zone, which is commonly defined as the region in which liquid water can exist on the surface of the planet (HZ, Leconte et al. 2013; Kopparapu et al. 2013, 2014, 2016). This motivates us in studying an Archean-like atmosphere for a planet orbiting around M-dwarf. However, M-dwarfs emit more radiation in the near-infrared (NIR) regime but less in the ultraviolet (UV) regime when compared to G-dwarfs such as the Sun. Additionally, M-dwarfs are smaller, and therefore fainter, than G-dwarfs. As a result, the orbital period of the planets in the HZ is likely to be synchronised with their rotation period (Dole 1964; Kasting et al. 1993; Guillot et al. 1996; Barnes 2017), meaning that they have a permanent dayside and nightside. These two factors have important implications for the climate of planets orbiting such stars (Joshi et al. 1997).

Firstly, the shift of stellar radiation to longer wavelengths when moving from a G-dwarf to an M-dwarf results in higher absorption by, for example, water vapor H$_2$O and methane CH$_4$, in the planetary atmosphere. This leads to enhanced heating in the upper atmosphere and thus a greater stability against convection (Eager-Nash et al. 2020) and an increase in overall cloud coverage for an M-dwarf planet. Secondly, the circulation of tidally-locked planets is also driven by the large day-night contrast between the two hemispheres (Showman et al. 2013; Pierrehumbert & Hammond 2019; Zhang 2020), which can be broken down into an overturning circulation (Showman et al. 2013), stationary waves (Sardeshmukh & Hoskins 1988; Showman & Polvani 2010), and a super-rotating equatorial jet component (Showman & Polvani 2010, 2011; Tsai et al. 2014; Hammond & Pierrehumbert 2018). Hammond & Lewis (2021) found that the divergent overturning circulation dominates the day-to-night heat transport in these tidally-locked planets. Sergeev et al. (2022b) found that the simulated circulation on the tidally-locked TRAPPIST-1e could be prone to bistability exhibiting either a single equatorial prograde jet or dual mid-latitude prograde jets. This however depends on the model setup, physical parameterisation of convection and cloud radiative effect.

Next generation telescopes such as the James Webb Space Telescope (JWST), the Extremely Large Telescope (ELT), and the Giant Magellan Telescope (GMT) enable studies directed towards characterising these potentially habitable planets orbiting around M-dwarfs (Snellen et al. 2015; Gillon et al. 2017; Fauchez et al. 2019; Gillon et al. 2020; Turbet et al. 2020a). Due to its proximity (located 12 pc away) and high frequency of transits, TRAPPIST-1e is a popular candidate for future atmospheric characterisation (Gillon et al. 2016, 2017; Turbet et al. 2018; Fauchez et al. 2019, 2022; Initiative et al. 2023). This motivates climate modelling studies for a variety of at-mospheric scenarios (e.g. Turbet et al. 2018, 2022; Sergeev et al. 2022a).

Turbet et al. (2018) found that when the initial CO$_2$ content of the simulated atmosphere of TRAPPIST-1e is above a certain value, which is dependent on the background N$_2$, it is stable against CO$_2$ condensation due to its greenhouse effect and heat redistribution. Turbet et al. (2018) also explored the impact of Titan-like atmospheric compositions for the TRAPPIST-1 planets, with CH$_4$ surface partial pressures between 10 Pa and 10 MPa. Turbet et al. (2018) found that increasing CH$_4$ partial pressure initially led to a global surface cooling due to radiative cooling by stratospheric CH$_4$, similar to the results of Pierrehumbert (2010). Yet, Turbet et al. (2018) further showed that with enough CH$_4$, the cooling can be balanced, or even stopped by strong warming due to its tropospheric greenhouse effect. However there have been limited studies exploring the effect of haze for such a planet. We have therefore chosen TRAPPIST-1e to be our target planet, when studying the atmospheric dynamics under various CO$_2$ and CH$_4$ gas concentrations and haze profiles.

This work builds upon the work of Eager-Nash et al. (2023), which studied the effect of varying CH$_4$ and CO$_2$ partial pressures on the Archean Earth in a 3D GCM, and Mak et al. (2023), which explored the effect of prescribing various haze profiles by adopting the same Archean model setup. Note that both of these studies focused on a rapidly rotating planet around a Sun-like star, while in this work we focus on a tidally-locked planet around an M-dwarf. In this study we use the UM to perform 3D simulations of the climate of tidally-locked TRAPPIST-1e across a range of values of pCH$_4$ and pCO$_2$, with and without haze. The haze is prescribed to be horizontally uniform and is assumed to be representative of the Archean conditions. It is generated using the Atmos 1D photochemical model (as used by Arney et al. 2016, 2017; Mak et al. 2023). Although the focus of this work is TRAPPIST-1e, the results could be applicable to other tidally-locked planets. The rest of this paper is structured as follows. The photochemical model, radiative transfer model and climate model configurations are described in Sec. 2, along with the model parameter space. The results are presented in Sec. 3, followed by the discussion and future work in Sec. 4. Afterwards, our conclusions are presented in Sec. 5.

## 2 MODELS

Since this work is based on Mak et al. (2023), we summarise the model setup here and note that the detailed model description can be found in Sec. 2 of Mak et al. (2023). We simulate photochemical haze production using the 1D photochemical model Atmos, with the model configuration described in Sec. 2.1. Afterwards, we calculate the optical properties with the radiative transfer code described in Sec. 2.2 and prescribe the haze and varying gas profiles within the 3D GCM, the UM, which is explained in Sec. 2.3.

### 2.1 Photochemical Model: Atmos

We adopt the 1D photochemistry module from the Atmos model[2] to generate Archean-like haze to be prescribed in the climate model. The Archean chemistry scheme is based on the model described in Arney et al. (2016), with the model updates described in Lincowski et al. (2018), Teal et al. (2022) and Mak et al. (2023), and the boundary conditions described in Teal et al. (2022). The model comprises of

---
[2] https://github.com/VirtualPlanetaryLaboratory/atmos

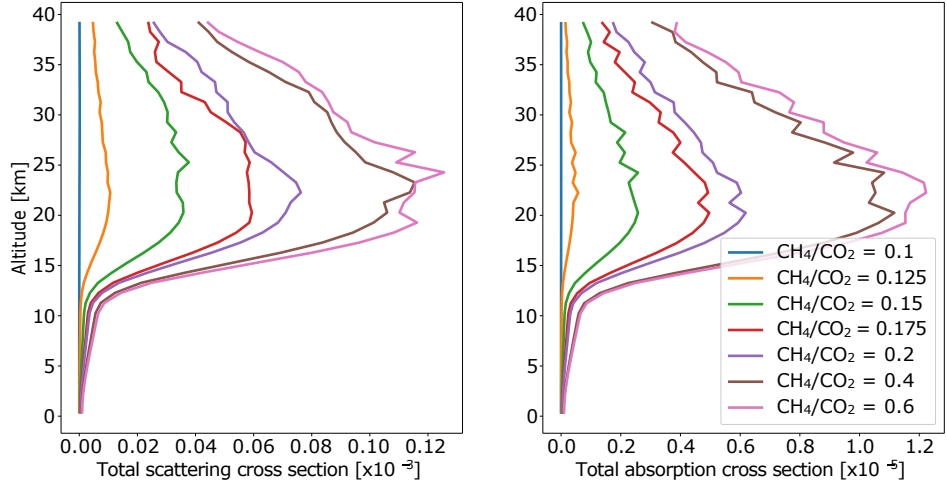

**Figure 1.** Total scattering (left) and absorption (right) cross section of the haze profiles plotted at the peak wavelength ∼1 $\mu$m of the BT-Settl model (Rajpurohit et al. 2013).

**Table 1.** The combination of pCH$_4$ and pCO$_2$ adopted in the simulations. Simulations that are not performed are denoted with "–".

| CH$_4$/CO$_2$ | 0.02 | 0.04 | 0.06 | 0.08 | 0.1 | 0.125 | 0.15 | 0.175 | 0.2 | 0.4 | 0.6 |
|---|---|---|---|---|---|---|---|---|---|---|---|
| pCO$_2$ [Pa] | | | | | pCH$_4$ [Pa] | | | | | | |
| 100 | – | – | – | – | 10 | 12.5 | 15 | 17.5 | 20 | 40 | 60 |
| 1000 | – | – | – | – | 100 | 125 | 150 | 175 | 200 | 400 | 600 |
| 3,000 | – | – | – | – | 300 | 375 | 450 | 525 | 600 | 1,200 | 1,800 |
| 5,000 | – | – | – | – | 500 | 625 | 750 | 875 | 1,000 | 2,000 | 3,000 |
| 10,000 | 200 | 400 | 600 | 800 | 1,000 | 1,250 | 1,500 | 1,750 | 2,000 | – | – |
| 20,000 | 400 | 800 | 1,200 | 1,600 | 2,000 | 2,500 | 3,000 | 3,500 | – | – | – |

**Table 2.** The maximum zonal wind speed [m s$^{-1}$] of the jet structure of all simulations. The corresponding value of $\sigma$ is included in brackets. The wind structure of each simulation is also denoted, with the single jet structure denoted as "SJ" and the double jet structure denoted as "DJ".

| pCO$_2$ [Pa] | CH$_4$/CO$_2$ | | | | | |
|---|---|---|---|---|---|---|
| | 0 | 0.02 | 0.04 | 0.06 | 0.08 | 0.1 |
| 100 (noAER) | DJ: 69.3 (0.33) | – | – | – | – | SJ: 44.9 (0.41) |
| 1000 (noAER) | DJ: 45.8 (0.41) | – | – | – | – | DJ: 82.4 (0.25) |
| 3,000 (noAER) | DJ: 45.7 (0.41) | – | – | – | – | DJ: 114.0 (0.072) |
| 3,000 (AER) | – | – | – | – | – | DJ: 64.3 (0.35) |
| 5,000 (noAER) | DJ: 46.9 (0.41) | – | – | – | – | DJ: 62.2 (0.34) |
| 10,000 (noAER) | DJ: 46.4 (0.40) | DJ: 71.0 (0.27) | DJ: 69.2 (0.30) | DJ: 64.4 (0.33) | DJ: 124.8 (0.041) | SJ: 42.6 (0.51) |
| 20,000 (noAER) | DJ: 47.3 (0.34) | SJ: 78.2 (0.19) | SJ: 41.1 (0.50) | SJ: 42.6 (0.49) | SJ: 41.8 (0.49) | SJ: 43.3 (0.53) |

| pCO$_2$ [Pa] | CH$_4$/CO$_2$ | | | | | |
|---|---|---|---|---|---|---|
| | 0.125 | 0.15 | 0.175 | 0.2 | 0.4 | 0.6 |
| 100 (noAER) | DJ: 49.4 (0.41) | SJ: 43.9 (0.42) | DJ: 124.7 (0.045) | SJ: 45.1 (0.42) | SJ: 42.8 (0.42) | DJ: 86.3 (0.14) |
| 1000 (noAER) | DJ: 82.0 (0.25) | DJ: 82.5 (0.22) | DJ: 124.7 (0.045) | DJ: 120.6 (0.058) | DJ: 55.9 (0.38) | DJ: 79.1 (0.14) |
| 3,000 (noAER) | DJ: 62.9 (0.35) | DJ: 58.8 (0.38) | DJ: 56.6 (0.38) | DJ: 78.4 (0.16) | DJ: 93.8 (0.10) | SJ: 42.7 (0.56) |
| 3,000 (AER) | DJ: 50.0 (0.34) | DJ: 71.9 (0.11) | SJ: 42.4 (0.54) | SJ: 41.4 (0.54) | DJ: 165.5 (0.015) | DJ: 164.2 (0.015) |
| 5,000 (noAER) | DJ: 80.7 (0.16) | DJ: 78.5 (0.16) | DJ: 77.1 (0.14) | DJ: 76.2 (0.14) | SJ: 42.7 (0.55) | SJ: 43.1 (0.55) |
| 10,000 (noAER) | SJ: 43.3 (0.51) | SJ: 41.9 (0.50) | SJ: 43.1 (0.55) | SJ: 43.1 (0.54) | – | – |
| 20,000 (noAER) | SJ: 43.0 (0.52) | SJ: 42.6 (0.52) | SJ: 42.5 (0.52) | – | – | – |

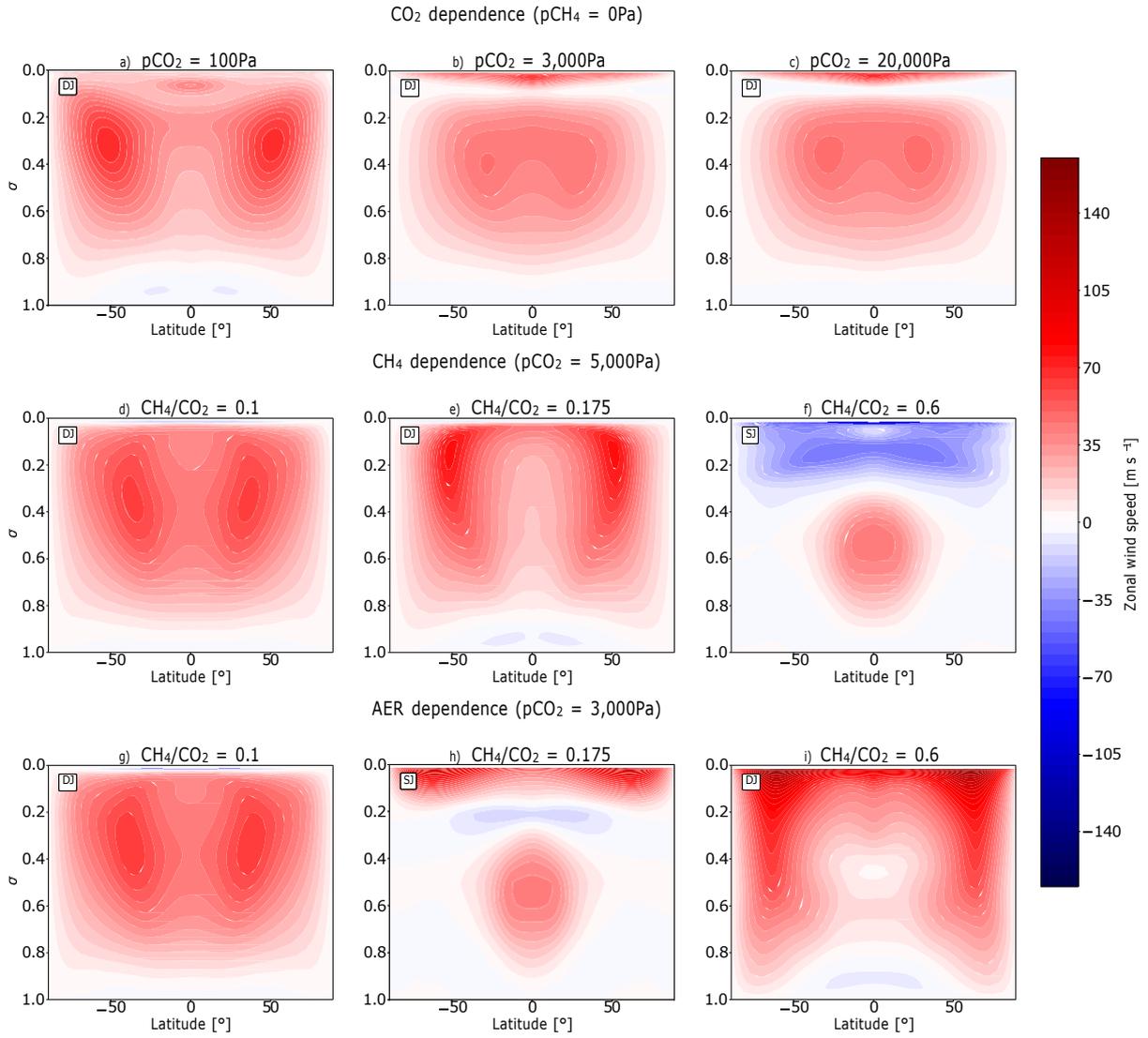

**Figure 2.** Zonal mean eastward wind distribution of the simulations. a, b and c shows the noAER simulations with pCO$_2$ = 100, 3,000 and 20,000 Pa (pCH$_4$ fixed at 0 Pa). d, e and f show the noAER simulations with CH$_4$/CO$_2$ = 0.1, 0.175 and 0.2 (pCO$_2$ fixed at 5,000 Pa). g, h and i show the AER simulations with CH$_4$/CO$_2$ = 0.1, 0.175 and 0.2 (pCO$_2$ fixed at 3,000 Pa). $\sigma$ (the y-axis) is the air pressure divided by the pressure at the surface. The wind structure of each simulation is also denoted, with the single jet structure denoted as "SJ" and the double jet structure denoted as "DJ".

433 chemical reactions and 76 chemical species. Hydrocarbon haze precursors are formed through the reactions:

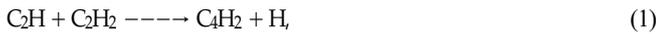
$$C_2H + C_2H_2 \dashrightarrow C_4H_2 + H, \tag{1}$$

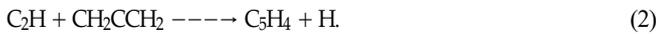
$$C_2H + CH_2CCH_2 \dashrightarrow C_5H_4 + H. \tag{2}$$

The full haze formation chemical process is not well understood despite both laboratory and theoretical studies (Hallquist et al. 2009; Hicks et al. 2015). However, in this work, the haze precursors C$_4$H$_2$ and C$_5$H$_4$ are assumed to form haze particles immediately after formation (Lavvas et al. 2008; Hörst 2017). The haze particles have a mass density of 0.64 g cm$^{-3}$ (Pavlov et al. 2001; Arney et al. 2016; Mak et al. 2023). They are assumed to have a log-normal distribution in radius, with a geometric standard deviation of 1.5. As described in Sec. 1, the haze profiles used for all the simulations in this work are the same set as used in Mak et al. (2023), where it is produced for a fixed pCO$_2$ of 3,000 Pa but for varying pCH$_4$ values such that the CH$_4$/CO$_2$ ratio is 0.1, 0.125, 0.15, 0.175, 0.2, 0.4 and 0.6. Note that this photochemical haze is Archean-like, i.e. generated from the solar spectrum of the Sun at 2.9 Ga (Claire et al. 2012) rather than from the stellar spectrum of an M-dwarf (Peacock et al. 2019; Wilson et al. 2021). We understand that different M-dwarf spectra can drive different outcomes for haze formation at a given CH$_4$/CO$_2$ ratio (Arney et al. 2017). However, the motivation behind this choice is that we would like to study the effect of haze on the dynamics of the tidally-locked planet, compared with its impact on a rapidly-rotating planet (Mak et al. 2023), rather than exploring the specific production of haze on TRAPPIST-1e. Additionally we do not have enough data to constrain the haze production environment for a planet such as TRAPPIST-1e. As a result we have opted the same haze profiles as in Mak et al. (2023, see their Fig. 1).

### 2.2 Radiative Transfer Model: SOCRATES

We use the "Suite Of Community RAdiative Transfer codes based on Edwards & Slingo (1996)" (SOCRATES), a two-stream radiative

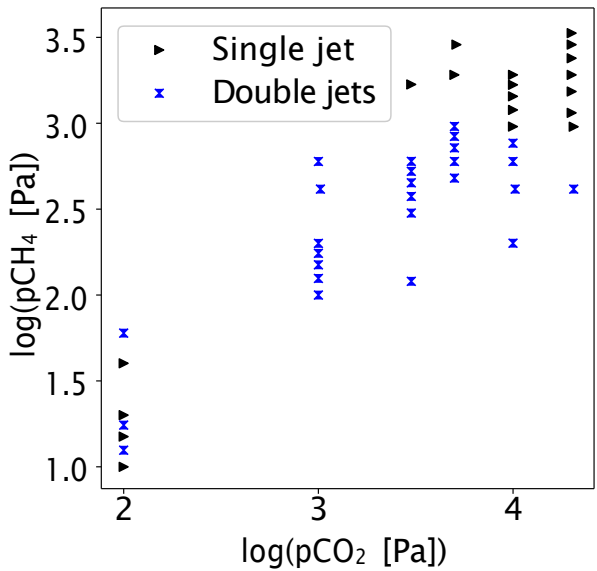

**Figure 3.** Distribution of noAER simulations with single jet or double jet structure.

transfer code, to calculate the optical properties of the photochemical haze particles by treating them as spherical particles and performing Mie scattering to obtain their scattering and absorption coefficients and asymmetry factors. Future work will update the model to include modified mean field theory of the fractal particles to calculate their optical properties (Botet et al. 1997; Tazaki & Tanaka 2018; Lodge et al. 2023) (detail discussed in Sec. 4.2).

Following Mak et al. (2023), the refractive indices are taken from Khare et al. (1984) and He et al. (2022). Khare et al. (1984) measured the optical properties for a wavelength range covering 0.027 $\mu$m to 920 $\mu$m by generating thin films of tholins by electric discharge of a mixture of 0.9 $N_2$ gas and 0.1 $CH_4$ gas. Similarly He et al. (2022) measured the optical properties of haze across the wavelength range 0.4 $\mu$m to 3.5 $\mu$m by using vacuum spectroscopy in a mixture of 5% of $CH_4$ in $N_2$. The real part of the refractive indices from He et al. (2022) share a similar magnitude to those from Khare et al. (1984). However the imaginary part from He et al. (2022) is almost half of that from Khare et al. (1984). The differences could be due to the different experimental conditions and the choices of optical model. For Mie scattering, we adopt the values provided by Khare et al. (1984) for the full wavelength range. However we adopt the updated values from He et al. (2022) instead in the wavelength regime of 0.4 $\mu$m to 3.5 $\mu$m.

The detail of the haze optical properties generated can be found in Mak et al. (2023). Since the optical properties of the haze are wavelength-dependent, the strength of their extinction effect is sensitive to the stellar spectrum used. We use the spectrum of an M-dwarf from the BT-Settl model (Rajpurohit et al. 2013) (which is the stellar spectrum used by the climate model, and is discussed in the next section), despite the fact that the haze is generated from an Archean solar spectrum as mentioned previously. The stellar parameters used in this paper follow Grimm et al. (2018), which are adopted in Sergeev et al. (2022a). The stellar spectrum is for an effective temperature of 2600 K, logg=5 and Fe/H = 0. The selected M-dwarf spectrum from the BT-Settl model (Rajpurohit et al. 2013) peaks at ∼1 $\mu$m, and is half of the peak intensity of the solar spectrum of the Sun at 2.9 Ga (Claire et al. 2012), which is at ∼0.45 $\mu$m. The total scattering and absorption cross sections of the haze are shown in Fig. 1, plotted at the peak wavelength ∼1 $\mu$m of the BT-Settl model. Compared to when irradiated at the peak wavelength ∼0.45 $\mu$m of the Archean solar spectrum (see Fig. 2, Mak et al. 2023), the total scattering cross section in this work is roughly halved, with a slightly smaller total absorption cross section as well. This is due to a smaller stellar flux from a cooler M-dwarf (Rajpurohit et al. 2013) compared to the hotter G-dwarf, our Sun, at 2.9 Ga (Claire et al. 2012), leading to a weaker extinction effect from the haze. Note that Ridgway et al. (2023) have constructed an M-dwarf stellar spectrum by combining data from the MUSCLES survey (France et al. 2016; Loyd et al. 2016; Youngblood et al. 2016) and the Proxima Centauri stellar spectrum from Ribas et al. (2017) and they found that it had a larger far UV (FUV) flux compared to the equivalent BT-Settl model. This would alter the photochemistry in the planetary atmosphere and change the haze production, which is not captured in our work here. However, the extinction effect of haze should remain similar. This is because the optical properties of the haze are largely dependent on the intensity of the flux and both stellar spectra from Rajpurohit et al. (2013) and Ridgway et al. (2023) peak in the NIR range.

To apply SOCRATES in the UM, configuration files (also referred to as "spectral files") are constructed, which parameterise the optical properties of gases, clouds and haze in wavelength bands in the shortwave and longwave part of the spectrum (Amundsen et al. 2014; Lines et al. 2018; Manners et al. 2022). The spectral files are almost identical to the one used in Mak et al. (2023), except the fact that here, the stellar spectrum is taken from the BT-Settl model (Rajpurohit et al. 2013). The shortwave configuration files cover the wavelength range of 0.2 $\mu$m to 20 $\mu$m of the stellar spectrum and is binned into 43 bands. The longwave configuration files cover the wavelength range of 3.33 $\mu$m to 10 mm of the thermal radiation from the planet and the atmosphere and is binned into 17 bands. These files also contain the optical properties of aerosols and radiatively active gas species. In each wavelength band, the gaseous absorption is calculated using the correlated-$k$ method. $k$-terms are generated using absorption line lists from HITRAN 2012 (Rothman et al. 2013). Collision-induced absorption from the gases are included, covering $N_2$-$CH_4$, $N_2$-$N_2$, $CO_2$-$CO_2$ from HITRAN (Karman et al. 2019), and $CH_4$-$CO_2$ (Turbet et al. 2020b).

### 2.3 Climate Model: the Unified Model

We use the Unified Model (UM), the 3D General Circulation Model developed by the Met office, to simulate the climate evolution of TRAPPIST-1e. The UM has been used to study modern Earth (Walters et al. 2019; Andrews et al. 2020) as well as the Archean Earth (Eager-Nash et al. 2023; Mak et al. 2023), other planets in the solar system such as Mars (McCulloch et al. 2023), terrestrial exoplanets (Boutle et al. 2020; Sergeev et al. 2022b; Ridgway et al. 2023) and gas giant exoplanets (Mayne et al. 2019; Christie et al. 2021; Zamyatina et al. 2023). The dynamical core of the UM, ENDGame (Even Newer Dynamics for General atmospheric modelling of the environment), solves the non-hydrostatic, deep-atmosphere equations of motion in the atmosphere with varying gravity (see for discussion Wood et al. 2014). Regarding the planetary parameters, the semi-major axis is taken to be 0.02928 AU. Both the orbital and rotation period are taken to be 6.1 Earth days. There is no obliquity and eccentricity. The gravity is set to be 9.12 m s$^{-2}$. Note that the model setup follows Eager-Nash et al. (2023) and Mak et al. (2023), except for the fact that the slab ocean depth is lower in our simulations here compared to their Archean simulations. This would only affect how quickly our simulations reach equilibrium and has no effect on our results. To summarise, a horizontal grid spacing of 2.5° in longitude and 2° in latitude are used in this study. The vertical grid is quadratically stretched into 38 layers with the top level at 39.25 km. This allows a higher resolution near the surface. We define a simulation to be in steady state when the surface temperature fluctuation per Earth year is less than 0.5 K. It is then run for another standard length of 20 Earth years and the results are temporally averaged (mean) and analysed.

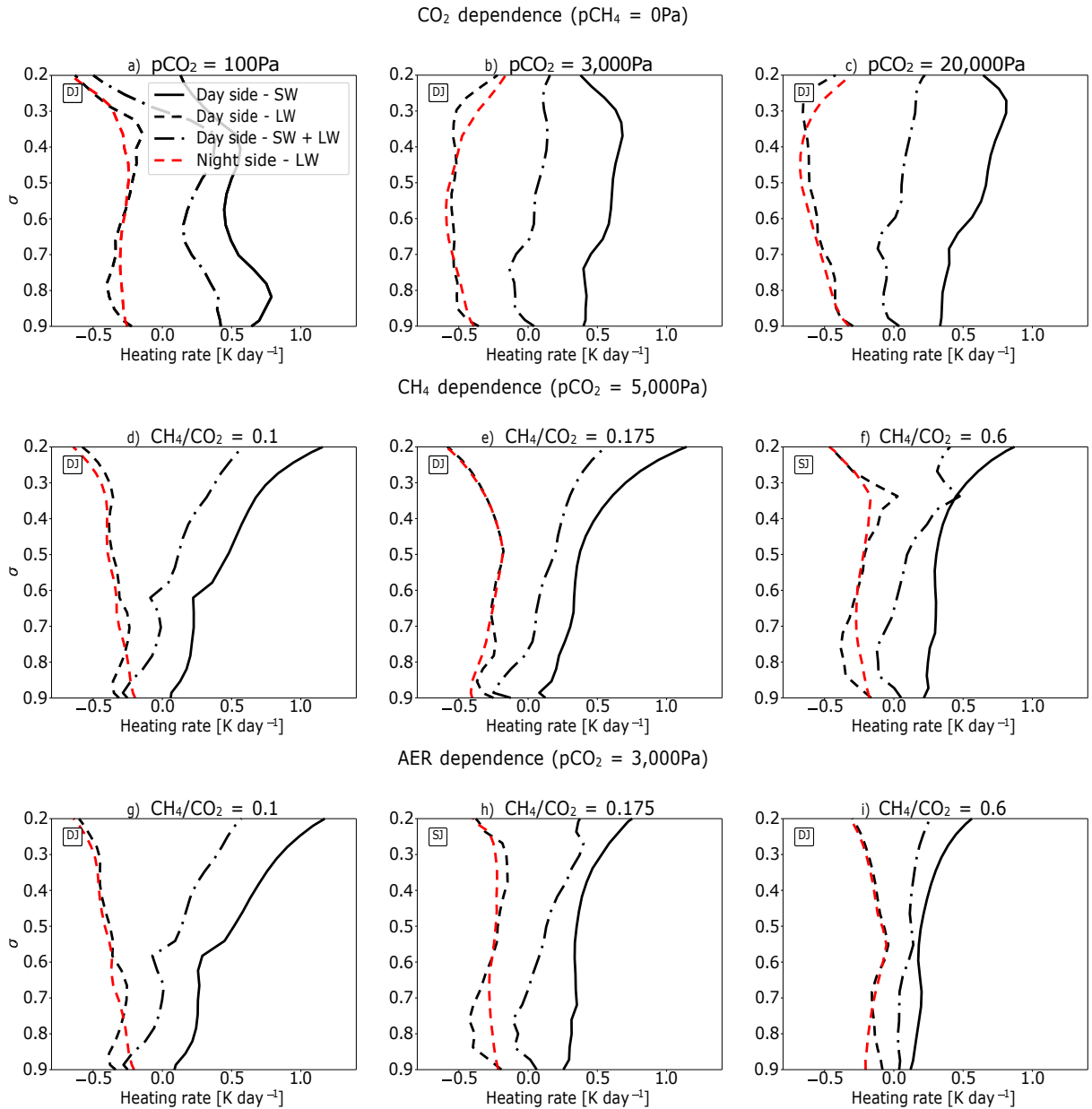

**Figure 4.** Averaged shortwave and longwave heating rates of the simulations. a, b and c show the noAER simulations with pCO$_2$ = 100, 3,000 and 20,000 Pa (pCO$_2$ fixed at 0 Pa). d, e and f show the noAER simulations with CH$_4$/CO$_2$ = 0.1, 0.175 and 0.2 (pCO$_2$ fixed at 5,000 Pa). g, h and i show the AER simulations with CH$_4$/CO$_2$ = 0.1, 0.175 and 0.2 (pCO$_2$ fixed at 3,000 Pa). $\sigma$ (the y-axis) is the air pressure divided by the pressure at the surface. The wind structure of each simulation is also denoted, with the single jet structure denoted as "SJ" and the double jet structure denoted as "DJ".

The atmosphere in our simulations comprises N$_2$, CO$_2$ and CH$_4$. The partial pressures of CH$_4$ and CO$_2$ are given in Tab. 1. Simulations with larger values of pCH$_4$ and pCO$_2$ are omitted as this gas mixture is beyond the limits of our radiative input data. The haze layers are prescribed based on the results of the photochemical model, meaning that they are fixed in time and horizontally uniform. It is possible that feedback from the climate conditions on the haze production could have a non-negligible impact. However, as we work towards a fully interactive, or prognostic, haze treatment, haze prescription is an important first step in understanding the implications of haze for planets such as TRAPPIST-1e. For brevity, in Sec. 3 where the results with and without hydrocarbon hazes are compared, simulations with haze are labelled "AER" (AER for AERosol), and those without haze "noAER" (noAER for no AERosol), following the labelling conventions from Arney et al. (2016) and Mak et al. (2023). For more intuitive interpretation, haze is also referred as "thin" when the CH$_4$/CO$_2$ ratio is small with weak optical impacts, and "thick" when the CH$_4$/CO$_2$ ratio is large with corresponding strong optical impacts.

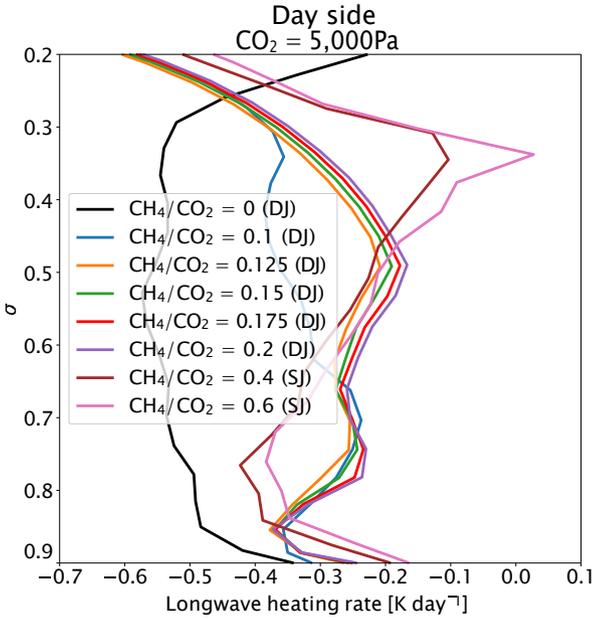

**Figure 5.** Averaged longwave heating rate on the dayside of the noAER simulation with varying $pCH_4$ ($pCO_2$ is fixed at 5,000 Pa). $\sigma$ (the y-axis) is the air pressure divided by the pressure at the surface. The wind structure of each simulation is also denoted, with the single jet structure denoted as "SJ" and the double jet structure denoted as "DJ".

## 3 RESULTS

We find that both gases and the haze share similar radiative mechanisms which affect the atmospheric structure of the tidally-locked planet in different ways. We therefore present our results into two different sections, each focusing on the dayside and the nightside of the planet, respectively. In each section, we discuss the similarities and differences in the impact on the atmospheric structure due to the presence of $CO_2$, $CH_4$ and the haze.

### 3.1 Dayside

Note that for a tidally-locked planet, only one side of the planet is permanently irradiated by its parent star. This day-to-night temperature contrast gives rise to the formation of jets, alongside the overturning circulation, which facilitates energy and atmospheric constituents transport from the dayside to the nightside of the planet (Hammond & Lewis 2021). Therefore to understand the atmospheric structure and how its key elements, such as clouds, are distributed, we first focus on the wind structure. In this paper, we focus on the tropospheric circulation. In all cases it exhibits superrotation. In general, there are two types of jet structure in our simulations: a single jet and a double jet, following the nomenclature of Sergeev et al. (2022b). Here we define a single jet to be a zonal mean zonal wind structure with only one peak with respect to the latitude. A double jet regime, on the other hand, has two peaks. Tab. 2 summaries the jet structure in each simulation, along with its maximum wind speed in the jet and their corresponding altitude in the $\sigma$ coordinate. $\sigma$ is the air pressure divided by the pressure at the surface. Fig. 2 shows the zonal wind distribution for noAER simulations with varying $pCO_2$ ($pCH_4$ fixed at 0 Pa), varying $pCH_4$ ($pCO_2$ fixed at 5,000 Pa), and those from AER simulations ($pCO_2$ fixed at 3,000 Pa). Simulations with other combinations of fixed $pCO_2$ and varying $pCH_4$ show similar results. We selected the most illustrative simulations. The single jet structure can be seen clearly in the form of a prograde jet in the equatorial region occurring in the lower atmosphere, for example as seen from Fig. 2f and h. The double jet structure can be seen clearly in the form of two prograde mid-latitude jets peaking in the upper atmosphere, for example as seen from Fig. 2a, e and i. There are also simulations exhibiting a transitional state between the two regimes, where there are two peaks with very similar magnitude appearing in the lower atmosphere, for example as seen in Fig. 2b and c. For conciseness, we still define these cases as having a double jet structure. However, this choice does not affect the conclusion of our finding.

Fig. 3 shows the distribution of noAER simulations with a single jet and a double jet structure. When $pCO_2$ is low, there is a mixture of both single jet and double jet structures in our simulations regardless of the value of $pCH_4$. However, when both $pCO_2$ and $pCH_4$ are high, the wind structure in the simulations tends to be a single jet. Fig. 4 shows the averaged shortwave and longwave heating rate in the corresponding simulations from Fig. 2. Fig. 4 only captures the altitude between $\sim 0.2<\sigma<0.9$ as this region shows the most drastic change in the shortwave and longwave heating rate. Fig. 4 shows that for the simulations that exhibit a single jet structure under high concentration of $CO_2$ and $CH_4$, they present, in general, a decrease of longwave cooling rate with altitude between $\sim 0.35<\sigma<0.75$, as seen from Fig. 4f and h. In addition the specific set of simulations with varying $pCH_4$ and when $pCO_2$ is fixed at 5,000 Pa, are shown in Fig. 5. Among the simulations presented in Fig. 5, only those where $CH_4/CO_2 = 0.4$ and 0.6 exhibit a single jet structure. They are also the only simulations that, in general, present a positive gradient with altitude in the longwave cooling rate of the atmosphere (between $\sim 0.35<\sigma<0.75$). In other words, from Fig. 4 and Fig. 5, the change in opacity due to different $CO_2$, $CH_4$ and haze concentrations changes the radiative heating in different parts of the atmosphere, therefore giving rise to various jet structure. The transition between these two states, was found to be very sensitive to parameter and setup choices in Sergeev et al. (2022b). Clearly, from our results, it is also sensitive to the atmospheric composition. For this work which looks at the impact of $CO_2$, $CH_4$ and haze on the climate of TRAPPIST-1e, a full analysis of what drives the dynamical flow into either regime is beyond the scope of what we are studying. Therefore, we reserve a deeper analysis of the dynamical evolution for a future work.

Fig. 6 shows the vertical temperature profiles for the dayside in the noAER simulations with varying $pCO_2$ ($pCH_4$ fixed at 0 Pa), varying $pCH_4$ ($pCO_2$ fixed at 5,000 Pa) and those from AER simulations ($pCO_2$ fixed at 3,000 Pa). The entire atmospheric column is warmer with increasing $pCO_2$ due to the increasing absorption of longwave radiation by the $CO_2$ (Fig. 6a). The temperature decreases gradually with altitude. However, in the upper atmosphere at $\sigma \sim 0.1$, the simulations are experiencing heating due to the shortwave absorption from $CO_2$, reaching a maximum temperature of $\sim 215$ K. This leads to a thermal inversion in the upper atmosphere. $CH_4$ is a stronger shortwave absorber, compared to $CO_2$. Therefore the shortwave heating in the upper atmosphere is larger compared to the $CO_2$ case (Fig. 6b). The maximum temperature in the upper atmosphere is $\sim 270$ K in this particular set of simulations. When $CH_4/CO_2 \leq 0.2$, the shortwave heating peaks near the top-of-atmosphere. However when $CH_4/CO_2 = 0.4$ and 0.6, a thermal inversion starts at $\sigma \sim 0.3$. These two simulations also exhibit a single equatorial jet, while the other simulations exhibit a double jet structure. In simulations with other values of $pCO_2$ that are not plotted here, a similar trend is apparent. For those with a double jet structure, their dayside temperature profile decreases gradually with increasing in altitude, similar to Fig. 6b when $CH_4/CO_2 \leq 0.2$. Yet for those with a single jet structure, a sharp increase in temperature at $\sigma \sim 0.3$ is present, similar to

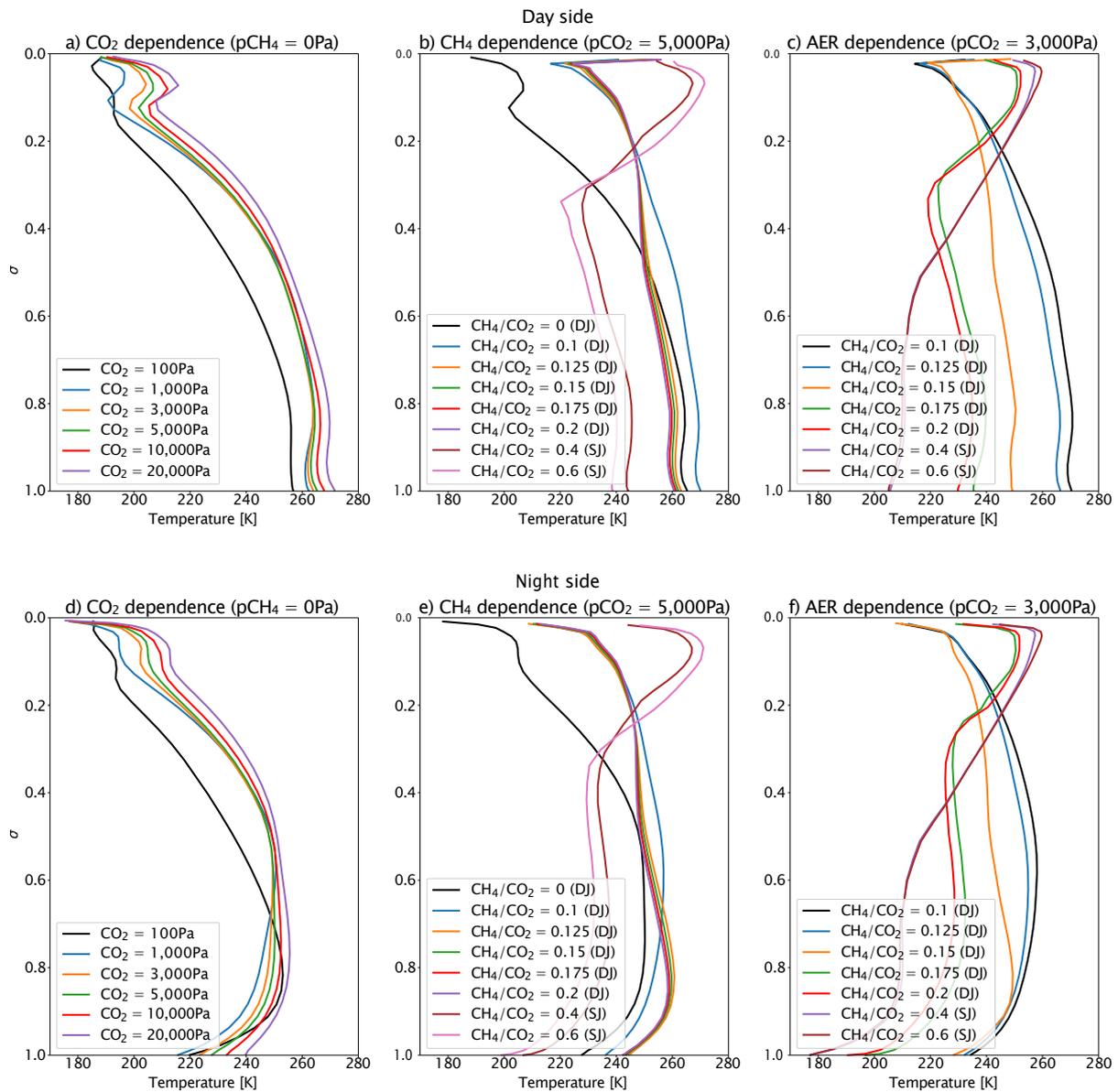

**Figure 6.** Temperature pressure profile from the dayside (top panel) and the nightside (bottom panel) of the simulations. a and d show the noAER simulations with varying $pCO_2$ (pCH$_4$ fixed at 0 Pa). b and e show the noAER simulations with varying $pCH_4$ ($pCO_2$ fixed at 5,000 Pa). c and f show the AER simulations (pCO$_2$ fixed at 3,000 Pa). $\sigma$ (the y-axis) is the air pressure divided by the pressure at the surface.

Fig. 6b when CH$_4$/CO$_2$ = 0.4 and 0.6. Furthermore, as it reaches the lower atmosphere, simulations with a larger pCH$_4$ result in a cooler lower atmosphere. This is due to the increasing shortwave absorption strength from increasing pCH$_4$, resulting in increasing stratospheric cooling and reduction of the intensity of shortwave radiation reaching the lower atmosphere. From Fig. 6c, under the presence of haze, this strong heating can also be seen in the upper atmosphere in each simulation, with the maximum temperature reaching ∼260 K. A thermal inversion is again apparent, when CH$_4$/CO$_2$ ≥ 0.175. Similar to the simulations for various CH$_4$ values, the temperature of the lower atmosphere reduces with an increasing thickness of the haze layer.

To explore the potential for habitability within our simulations, we study their surface temperature distributions. Fig. 7 shows the dayside mean surface temperature for noAER simulations with varying pCO$_2$ (pCH$_4$ fixed at 0 Pa), varying both pCH$_4$ and pCO$_2$, and those from AER simulations (pCO$_2$ fixed at 3,000 Pa) (the nightside profiles are discussed in Sec. 3.2). From Fig. 7a, the dayside mean surface temperature increases with increasing pCO$_2$, from ∼258.5 K when pCO$_2$ = 100 Pa, to ∼272.6 K when pCO$_2$ = 20,000 Pa (increase of ∼14.1 K). From Fig. 7b, increasing pCH$_4$ leads to the surface temperature first increasing and then decreasing. This peak occurs at a larger CH$_4$/CO$_2$ ratio when pCO$_2$ is small. For example, when pCO$_2$ = 3,000 Pa, this peak occurs when CH$_4$/CO$_2$ = 0.125. Yet when pCO$_2$ = 20,000 Pa, this peak occurs when CH$_4$/CO$_2$ = 0.02. Another peak is observed when pCO$_2$ = 100 Pa and CH$_4$/CO$_2$ = 0.175 but is likely to be caused by the change in circulation from a single to double jet structure compared to other simulations with the same pCO$_2$. This however does not affect the overall trend and the conclusions drawn here. From Fig. 7d, under the presence of a thin layer of haze when CH$_4$/CO$_2$ = 0.1, there is an increase of surface temperature of ∼4.9 K, compared to the corresponding noAER simulation. As the CH$_4$/CO$_2$ ratio increases and the haze layer becomes thicker, there is a sharp decrease in surface temperature, with values lower than that from the corresponding noAER simulations where CH$_4$/CO$_2$ > 0.1. When CH$_4$/CO$_2$ ≥ 0.4, the surface temperature reaches a plateau of ∼210 K. For all the simulations conducted, only the noAER simulations with high enough concentration of CO$_2$ and a low concentration of CH$_4$ have a dayside mean surface temperature above 273.15 K. Those sets of simulations are [pCO$_2$,pCH$_4$] = [10,000 Pa,400 Pa], [10,000 Pa,600 Pa], [10,000 Pa,800 Pa] and [20,000 Pa,800 Pa].

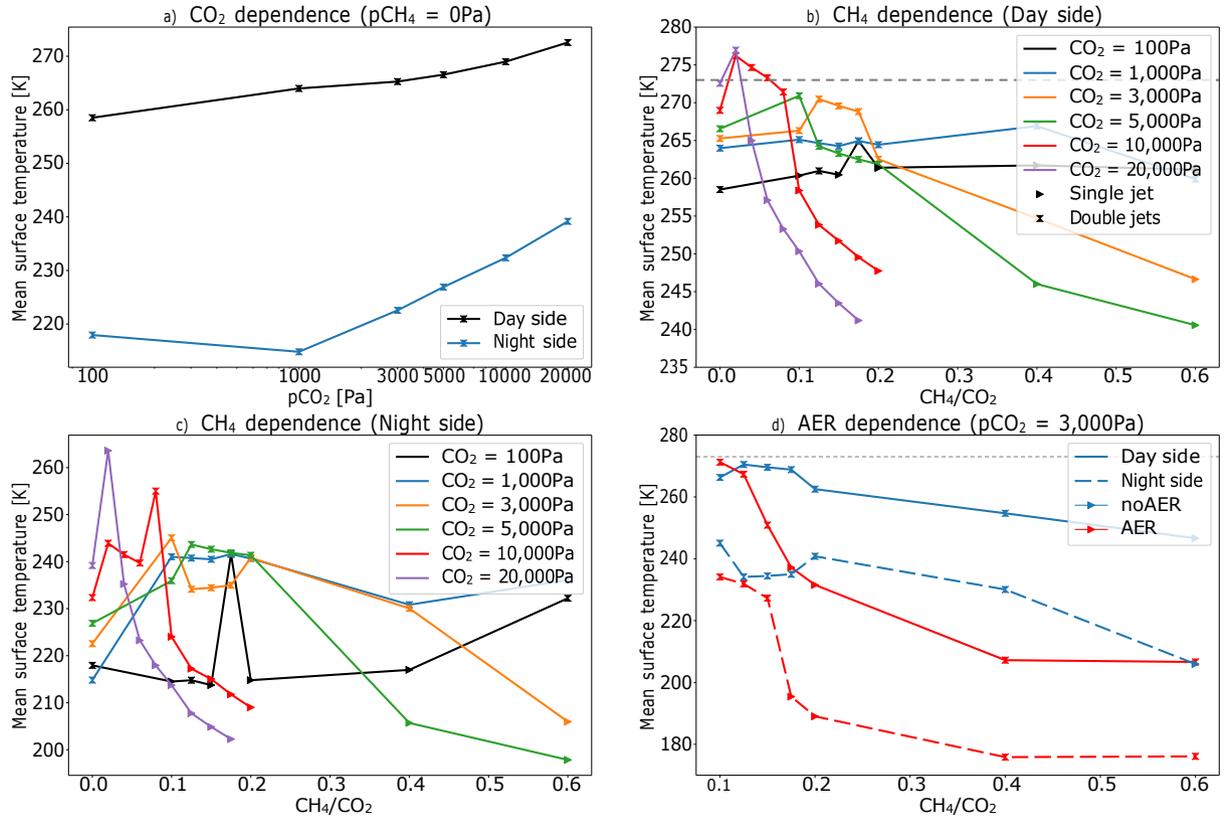

**Figure 7.** Dayside and nightside mean surface temperature of the simulations. a shows the noAER simulations with varying pCO$_2$ (pCH$_4$ fixed at 0 Pa). b and c show the noAER simulations with varying pCH$_4$ and pCO$_2$, respectively. d shows the AER simulations (pCO$_2$ fixed at 3,000 Pa). The water freezing point 273 K is marked as a habitability threshold. The wind structure of each simulation is also denoted, with the single jet structure denoted as "▶" and the double jet structure denoted as "✕".

Fig. 8 shows the surface temperature distribution on the entire planet for noAER simulations with varying pCO$_2$ (pCH$_4$ fixed at 0 Pa), varying pCH$_4$ (pCO$_2$ fixed at 20,000 Pa), and those from AER simulations. Their jet structure and the region where liquid water can be sustained, also referred to as the liquid-water region, are marked on the figure (except for Fig. 8i in which the surface temperature is below 273 K in every grid cell). Fig. 8 also shows arrows indicating the direction and magnitude of the surface wind. Simulations with other combinations of fixed pCO$_2$ and varying pCH$_4$ show similar results. We selected the most illustrative simulations. For simulations that exhibit a single jet structure, they have a more longitudinally narrow liquid-water region, for example as seen from Fig. 8e and f. On the contrary, for those that exhibit a double jet structure, the liquid-water region is more circular and covers a wider longitudinal range, for example as seen from Fig. 8c, g and h. Note that in Fig. 8d when pCO$_2$ = 20,000 Pa and when CH$_4$/CO$_2$ = 0.02, the liquid-water region extends to the longitude of ∼150° along the equator. This is driven by the near surface wind that results in a weaker advection of cold air from the east, i.e. from the night side of the planet. As a result, the near-surface the hot spot is effectively shifted eastward.

Since clouds also play an important role in determining the energy budget of the planet, it is vital to look at the cloud area distribution and its radiative effect. Fig. 9 shows the dayside cloud area fraction for noAER simulations with varying pCO$_2$ (pCH$_4$ fixed at 0 Pa), varying pCH$_4$ (pCO$_2$ fixed at 5,000 Pa), and those from AER simulations (pCO$_2$ fixed at 3,000 Pa). Simulations with other combinations of fixed pCO$_2$ and varying pCH$_4$ show similar results. We selected the most illustrative simulations. From Fig. 9a when pCO$_2$ is small, clouds concentrate in the polar and equatorial regions. From Fig. 9b and c, as pCO$_2$ increases, the cloud distribution becomes more even across latitudes but reduces in concentration. This is because when pCO$_2$ increases, the temperature in the atmosphere increases as well, as seen from Fig. 6a. This reduces the relative humidity in the atmosphere, therefore weakening cloud formation. Fig. 10 and Fig. 11 show the dayside top-of-atmosphere and surface cloud radiative effect, respectively, of the corresponding simulations to Fig. 9. The cloud radiative effect is calculated by deducting the net downward flux of the cloud-free simulation from that of the corresponding cloudy simulation. The calculation is performed on the level below the model top level to obtain the top-of-atmosphere cloud radiative effect. The calculation is then performed on the planetary surface to obtain the surface cloud radiative effect. From Fig. 10a and Fig. 11a, as pCO$_2$ increases, there is less upward shortwave at the top-of-atmosphere and surface, respectively, due to the reduced cloud concentration. Less longwave radiation is trapped in the atmosphere as well due to the decreased amount of clouds. From Fig. 9d, e and f, when pCH$_4$ increases, there are increasing cloud coverage near the surface and a more even distribution across latitudes. Sim-ilar trends can be seen in the AER case from Fig. 9g, h and i as the haze layer thickens. This is due to the reduction of temperature in the atmosphere, as seen from Fig. 6. This acts to increase the relative humidity, facilitating cloud formation. Yet due to the low temperature in the lower atmosphere, cloud tends to form near the surface instead, leading to a decrease of the cloud fraction in the upper atmosphere. The decreased cloud fraction in the upper atmosphere again reduces the intensity of reflected shortwave radiation at the top-of-atmosphere, as seen from Fig. 10b and c. This allows more shortwave radiation to reach the lower atmosphere and the surface. Therefore there is a reduced net upward shortwave flux on the surface, as seen from Fig. 11b and c. Similarly, since less longwave radiation is trapped in the atmosphere, a slight decrease can be seen in the downward longwave flux as seen from Fig. 10b, c, and Fig. 11b, and c.

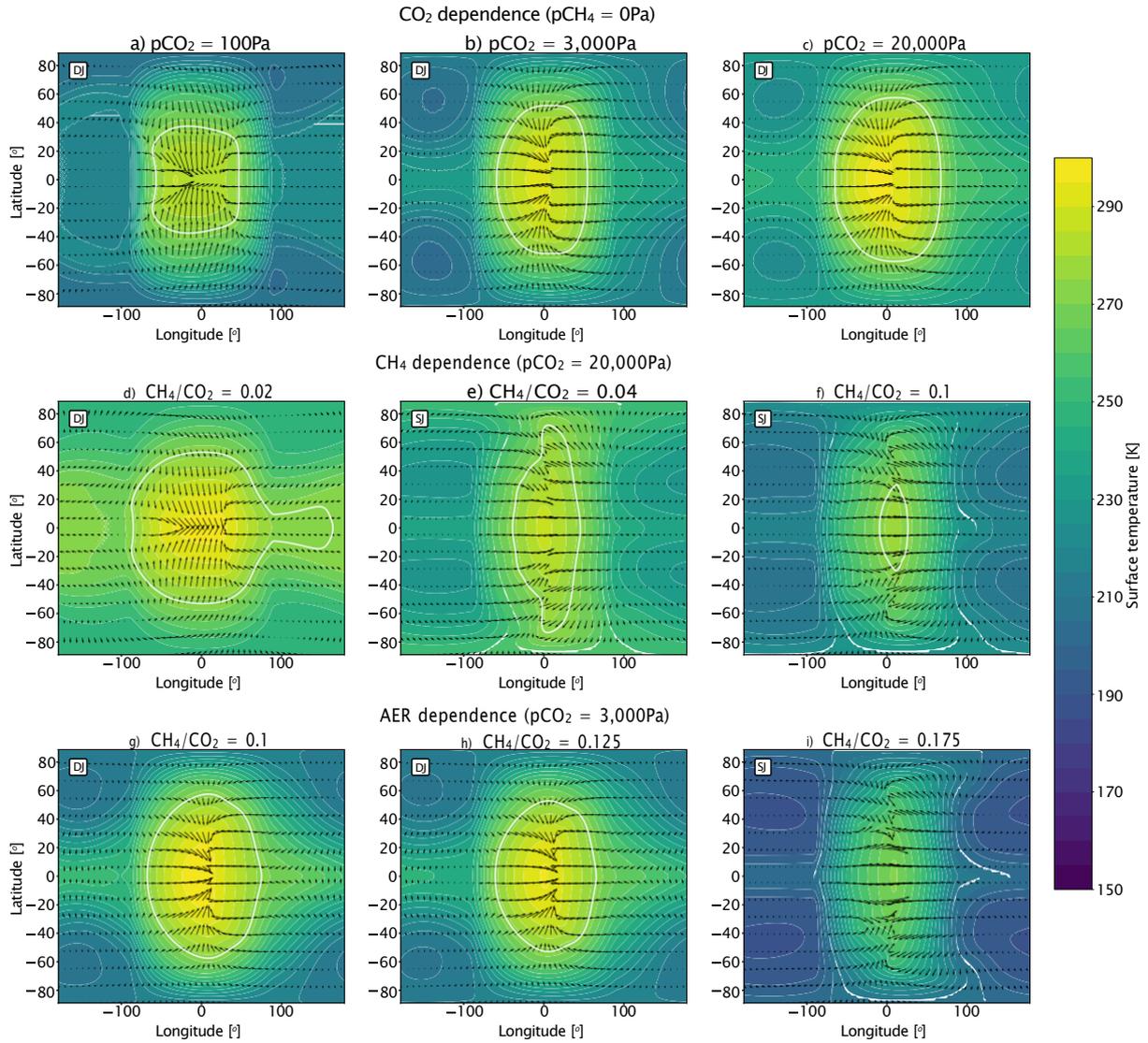

**Figure 8.** Contour plot of surface temperature of the simulations, overlapping with arrows indicating the direction and magnitude of the surface wind.. a, b and c show the noAER simulations with $pCO_2$ = 100, 3,000 and 20,000 Pa ($pCH_4$ fixed at 0 Pa). d, e and f show the noAER simulations with $CH_4/CO_2$ = 0.02, 0.04 and 0.1 ($pCO_2$ fixed at 20,000 Pa). g, h and i show the AER simulations with $CH_4/CO_2$ = 0.1, 0.125 and 0.175 ($pCO_2$ fixed at 3,000 Pa). The liquid-water region, defined as the area with the surface temperature larger than 273 K and can maintain liquid water on the surface, is denoted with a white line. The wind structure of each simulation is also denoted, with the single jet structure denoted as "SJ" and the double jet structure denoted as "DJ".

To summarise, in terms of the dayside of the tidally-locked planet, the longwave absorption from $CO_2$ becomes more significant as $pCO_2$ increases, leading to an increase in the atmospheric temperature. On the contrary, when $pCH_4$ increases or the haze layer thickens, their shortwave absorption strength becomes significant. This reduces the shortwave radiation reaching the lower atmosphere, therefore cooling down the entire lower atmosphere. Our simulations also exhibit different jet structures which are found to be sensitive to the atmospheric composition. For high values of $pCO_2$ and $pCH_4$, a single jet structure is preferred. The double jet structure produces a more circular liquid-water region, while the single structure produces a more longitudinally narrow region. In other words, the planet has a higher surface temperature when $pCO_2$ is large, $pCH_4$ is small, when they exhibit a double jet structure and when a thin haze layer is present.

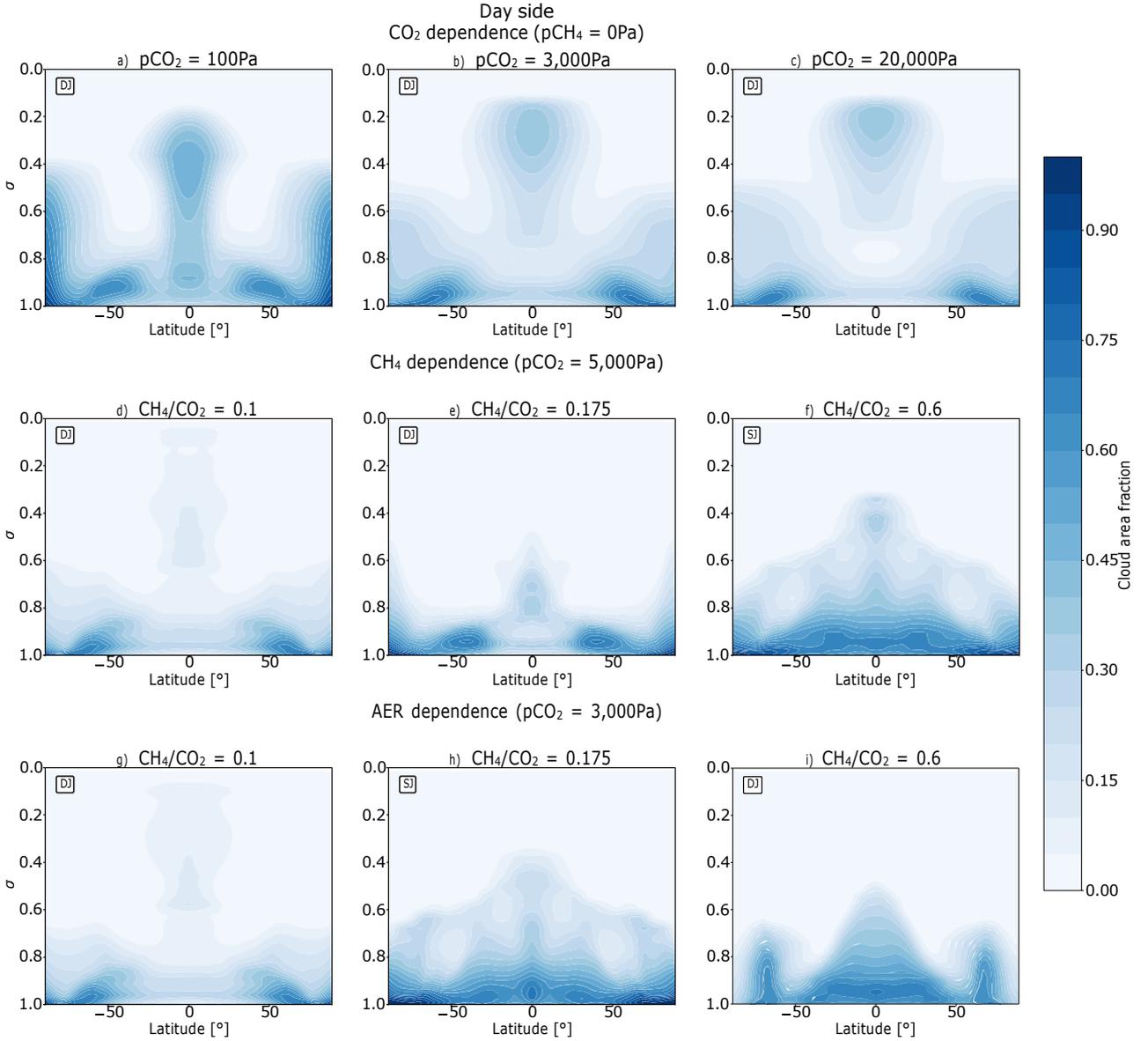

**Figure 9.** Dayside cloud area fraction of the simulations. a, b and c show the noAER simulations with pCO$_2$ = 100, 3,000 and 20,000 Pa (pCH$_4$ fixed at 0 Pa). d, e and f show the noAER simulations with CH$_4$/CO$_2$ = 0.1, 0.175 and 0.6 (pCO$_2$ fixed at 5,000 Pa). g, h and i show the AER simulations with CH$_4$/CO$_2$ = 0.1, 0.175 and 0.6 (pCO$_2$ fixed at 3,000 Pa). $\sigma$ (the y-axis) is the air pressure divided by the pressure at the surface.

## 3.2 Nightside

As discussed in Sec. 3.1, since the nightside of a tidally-locked planet is not irradiated by the star, its energy supply is mainly driven by the global wind circulation (Hammond & Lewis 2021). Here we analyse the night-side average temperature profiles (Fig. 6d-f). The temperature is higher at almost all altitudes for increasing pCO$_2$ owing to its increasing longwave absorption, similar to what occurs on the dayside (Fig. 6d). A thermal inversion is present near the surface, which is due to the jet transporting heat from the dayside to the nightside (Joshi et al. 2020). A weaker temperature gradient is seen with increasing pCO$_2$ and the simulations share similar temperature in the mid atmosphere at $\sigma \sim 0.5$ except when pCO$_2$ = 100 Pa. When pCO$_2$ = 100 Pa, the temperature inversion is at its strongest with a temperature difference of $\sim$34 K between $\sim 0.9 < \sigma < 1$. This is because this simulation, when pCO$_2$ = 100 Pa, exhibits the fastest zonal wind speeds for a circulation pattern with a double jet among the other simulations of the same set. Therefore its heat transport is the most efficient. When we consider only varying pCH$_4$ (Fig. 6e), increasing pCH$_4$ results in a warmer upper atmosphere for altitudes above the $\sigma \sim 0.2$ level. This is due to the dayside experiencing a warmer upper atmosphere at a similar altitude due to the shortwave absorption by CH$_4$, which is then transported to the nightside by the wind. This trend can also be seen in the AER simulations (Fig. 6f) when the haze layer is thickened. However between $\sigma \sim 0.2$ and the surface, the structure of atmosphere depends on the strength of the zonal wind in the simulation. For example, Fig. 6e shows that the simulation with CH$_4$/CO$_2$ = 0.125 has the warmest lower atmosphere as it has the fastest zonal wind speeds within a double jet regime. Similar to Fig. 6d, all simulations from Fig. 6e and f show thermal inversions near the surface due to the wind. However from Fig. 6e when CH$_4$/CO$_2$ = 0.4 and 0.6, and from Fig. 6f when CH$_4$/CO$_2 \geq$ 0.175, a second thermal inversion is seen above $\sigma \sim 0.3$. This comes from the thermal inversion at the same altitude from the dayside in the corresponding simulations, due to the shortwave heating from CH$_4$ and the haze, as discussed in Sec. 3.1.

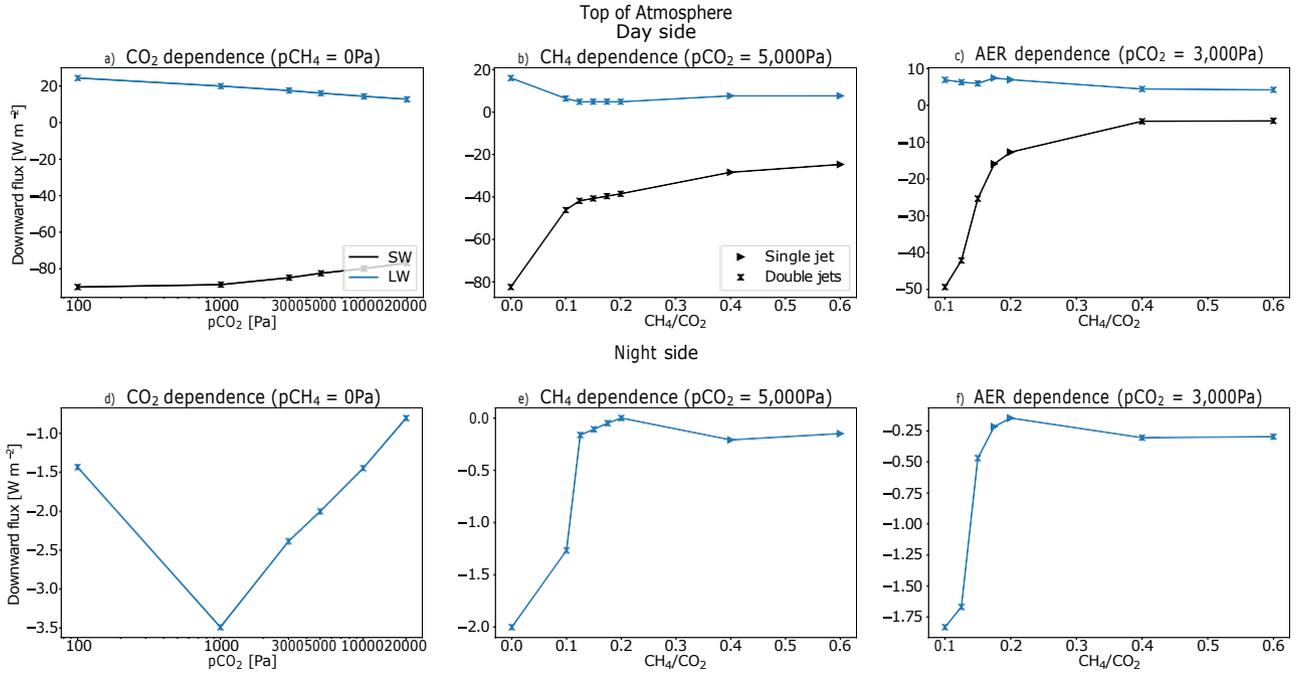

**Figure 10.** Top-of-atmosphere cloud radiative effect of the dayside (top panel) and the nightside (bottom panel) of the simulations. a and d shows the noAER simulations with varying pCO$_2$ (pCH$_4$ fixed at 0 Pa). b and e show the noAER simulations with varying pCH$_4$ (pCO$_2$ fixed at 5,000 Pa). c and f show the AER simulations with varying pCH$_4$ (pCO$_2$ is fixed at 3,000 Pa).

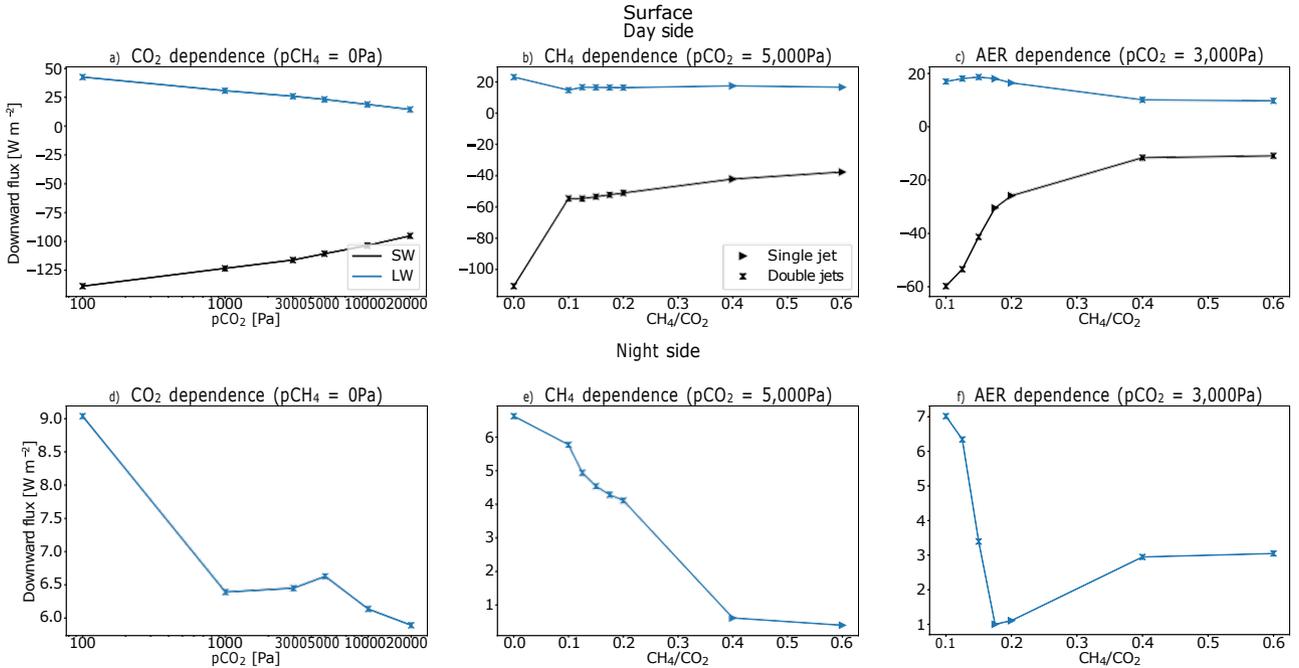

**Figure 11.** Surface cloud radiative effect of the dayside (top panel) and the nightside (bottom panel) of the simulations. a and d shows the noAER simulations with varying pCO$_2$ (pCH$_4$ fixed at 0 Pa). b and e show the noAER simulations with varying pCH$_4$ (pCO$_2$ fixed at 5,000 Pa). c and f show the AER simulations with varying pCH$_4$ (pCO$_2$ is fixed at 3,000 Pa).

Regarding the surface temperature on the nightside, none of our simulations show an averaged mean surface temperature above 273.15 K. From Fig. 7a, as we increase pCO$_2$ by fixing pCH$_4$ = 0 Pa, the mean surface temperature on the nightside increases by ∼21.2 K. The trend across our simulations, in general, follows that present on the dayside, except when pCO$_2$ = 1,000 Pa. This is due to its weaker double jet structure compared to when pCO$_2$ = 100 Pa. The heat is therefore transported less efficiently compared to the other simulations. From Fig. 7c, as we increase pCH$_4$, simulations with a double jet structure in general experience a warmer nightside. Furthermore, as discussed in Sec. 3.1, there is a peak of surface temperature on the dayside as we increase pCH$_4$. This peak can also be seen on the nightside, however at a different CH$_4$/CO$_2$ ratio. For example from Fig. 7b, when pCO$_2$ = 10,000 Pa, the temperature peaks when CH$_4$/CO$_2$ = 0.02 on the dayside. However from Fig. 7c on the nightside when pCO$_2$ = 10,000 Pa, the temperature peaks when CH$_4$/CO$_2$ = 0.08. This is due to the different jet structures and their strength variations. From Fig. 7d, again as discussed in Sec. 3.1, under the presence of haze, dayside warming of ∼4.9 K is observed only when CH$_4$/CO$_2$ = 0.1. The nightside, on the contrary, experiences a global cooling of ∼10.9 K when CH$_4$/CO$_2$ = 0.1. Again this is due to the strength difference of the double jet. From Fig. 12 and Tab. 2 when CH$_4$/CO$_2$ = 0.1, in the noAER simulation, the double jet peaks at ∼114.0 m s$^{-1}$. However in the AER simulation, the double jet peaks at ∼64.3 m s$^{-1}$. The nightside of the AER simulation therefore experiences a weaker heat transport, and hence a cooler surface temperature.

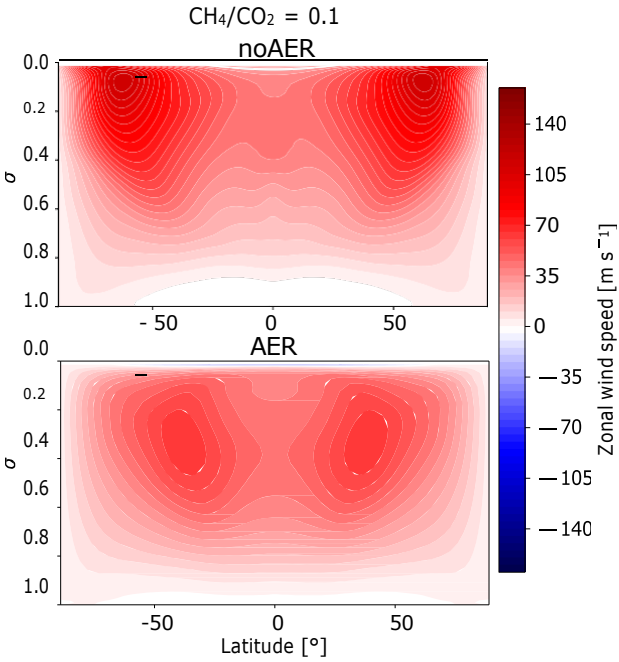

**Figure 12.** Zonal wind distribution of noAER and AER simulations when CH$_4$/CO$_2$ = 0.1 with pCO$_2$ fixed at 3,000 Pa. $\sigma$ (the y-axis) is the air pressure divided by the pressure at the surface.

Fig. 13 shows the nightside cloud area fraction for noAER simulations with varying pCO$_2$ (pCH$_4$ fixed at 0 Pa), varying pCH$_4$ (pCO$_2$ fixed at 5,000 Pa), and those from AER simulations (pCO$_2$ fixed at 3,000 Pa). Simulations with other combinations of fixed pCO$_2$ and varying pCH$_4$ show similar results. We selected the most illustrative simulations. From Fig. 13a, b and c, the cloud area fraction reduces as pCO$_2$ increases. This is due to the fact that the temperature is in general higher throughout the entire atmosphere, as seen from Fig. 6d. This reduces the relative humidity and therefore weakens the cloud formation. This behaviour resembles the trend seen on the dayside. Considering the top-of-atmosphere cloud radiative effect from Fig. 10d, the outgoing longwave radiation is reduced with increasing pCO$_2$. This is due to a thermal inversion on the nightside, as seen from Fig. 6d, with the planetary surface being colder than where the clouds are located. A reduction of cloud coverage means that a more significant part of the outgoing longwave radiation comes from the colder surface, leading to its overall decrease in intensity. The re-duction of cloud coverage also weakens the cloud greenhouse effect and reduces the intensity of longwave radiation reaching the surface, as seen in Fig. 11d. The noAER and AER simulations with varying pCH$_4$ share a similar behaviour here. When pCH$_4$ increases, both with and without haze, the cloud coverage is mostly limited to the lower atmosphere due to the thermal inversion locating closer to the surface, and the cold atmospheric structure. The absence of clouds in the upper atmosphere reduces the intensity of longwave radiation re-emitted back to space, as seen in Fig. 10e and f. This also weakens the cloud greenhouse effect in the atmosphere, as seen from Fig. 11e and f. However on the nightside for all simulations, a clear distinction in the cloud structure due to the single jet and double jet circulation structures is apparent. For those that exhibit a double jet structure, the cloud distribution concentrates in the equatorial and mid-to-high latitude regions and extends to the mid atmosphere. This can be seen in, for example Fig. 13d, g and i. For those that exhibit a single jet structure, their cloud distribution extends to the mid atmosphere in the equatorial region only and concentrates in the lower atmosphere close to the surface. This can be seen in, for example Fig. 13f and h. In summary, given the lack of incident radiation on the nightside, the atmospheric structure is dependent on the structure of the dayside and the circulation pattern from the dayside to the nightside. Therefore, understanding the nightside conditions requires an understanding of the circulation pattern for a given planet.

## 4 DISCUSSION

As discussed in Sec. 1 and Sec. 2, this paper builds on the work of Eager-Nash et al. (2023), who explored the impact of different CH$_4$ and CO$_2$ gas profiles on the simulated Archean Earth, and Mak et al. (2023), who used the same setup from Eager-Nash et al. (2023) but added prescribed spherical particle haze profiles (the same haze profiles used in this study). In the following sections, we compare our results of varying gas and haze profiles in the atmosphere of a tidally-locked planet to the results found for a rapidly rotating planet from Eager-Nash et al. (2023) and Mak et al. (2023), and from other previous work.

### 4.1 Rapidly Rotating Planet vs Tidally-locked Planet

#### 4.1.1 Effect of CH$_4$ and CO$_2$

Charnay et al. (2013) and Eager-Nash et al. (2023) found that increasing atmospheric pCO$_2$ will increase the temperature of the entire planet, in the case of rapid rotation. Similarly for a tidally-locked planet, our results show a warmer atmosphere on both dayside and nightside with increasing pCO$_2$ due to its greenhouse effect. There is also an increase of ∼14.1 K on the dayside when we increase pCO$_2$ from 100 Pa to 20,000 Pa. This is similar to what is found in Turbet et al. (2018).

Considering the effect of CH$_4$, Eager-Nash et al. (2023) found that as pCH$_4$ is increased in the rapidly rotating Archean Earth setup, the increase of global longwave radiative forcing initially dominates. Beyond the ratio of CH$_4$/CO$_2$ = 0.1, the total forcing is then dominated by the increase of negative global shortwave radiative forcing. This balance between the longwave warming and shortwave cooling, which has also been discussed in Byrne & Goldblatt (2015), leads to the surface temperature in their simulations peaking when CH$_4$/CO$_2$ = 0.1, reaching a warming of between ∼3.5 K to ∼7 K. A thermal inversion is also seen in their stratosphere due to the shortwave absorption from CH$_4$. In our work, the dayside also exhibits a similar change in surface temperature with $CH_4$ concentration. The surface temperature warming varies from ∼2.5 K to ∼11.4 K, across our simulations but at differing values of $CH_4/CO_2$ dependent on the value of $pCO_2$. For instance, in the noAER simulation when $pCO_2$ = 3,000 Pa, Eager-Nash et al. (2023) showed a maximum increase of surface temperature of ∼7 K when $pCH_4$ = 300 Pa ($CH_4/CO_2$ = 0.1). Yet here we find a maximum increase of surface temperature of ∼4.2 K when $pCH_4$ = 375 Pa on the dayside ($CH_4/CO_2$ = 0.125). $CH_4$ absorption is the most efficient in the infrared part of the spectrum and therefore the balance of longwave heating and shortwave cooling from $CH_4$ differs from what is found in Eager-Nash et al. (2023). Fig. 14 shows the net downward $CH_4$ radiative forcing at the planet's surface. Its shortwave radiative forcing becomes more negative when $pCH_4$ is increased, similar to what is found in Eager-Nash et al. (2023). Yet the longwave radiative forcing peaks when $CH_4/CO_2$ = 0.125, resulting in the total radiative forcing peaking at that ratio. However, similar to what is found in Eager-Nash et al. (2023), the surface temperature peaks at a smaller ratio when $pCO_2$ is high (i.e. at a larger value of $pCH_4$ and $pCO_2$). Similar to Eager-Nash et al. (2023), our work shows that as $pCH_4$ keeps increasing, the temperature drops due to shortwave absorption caused by $CH_4$.

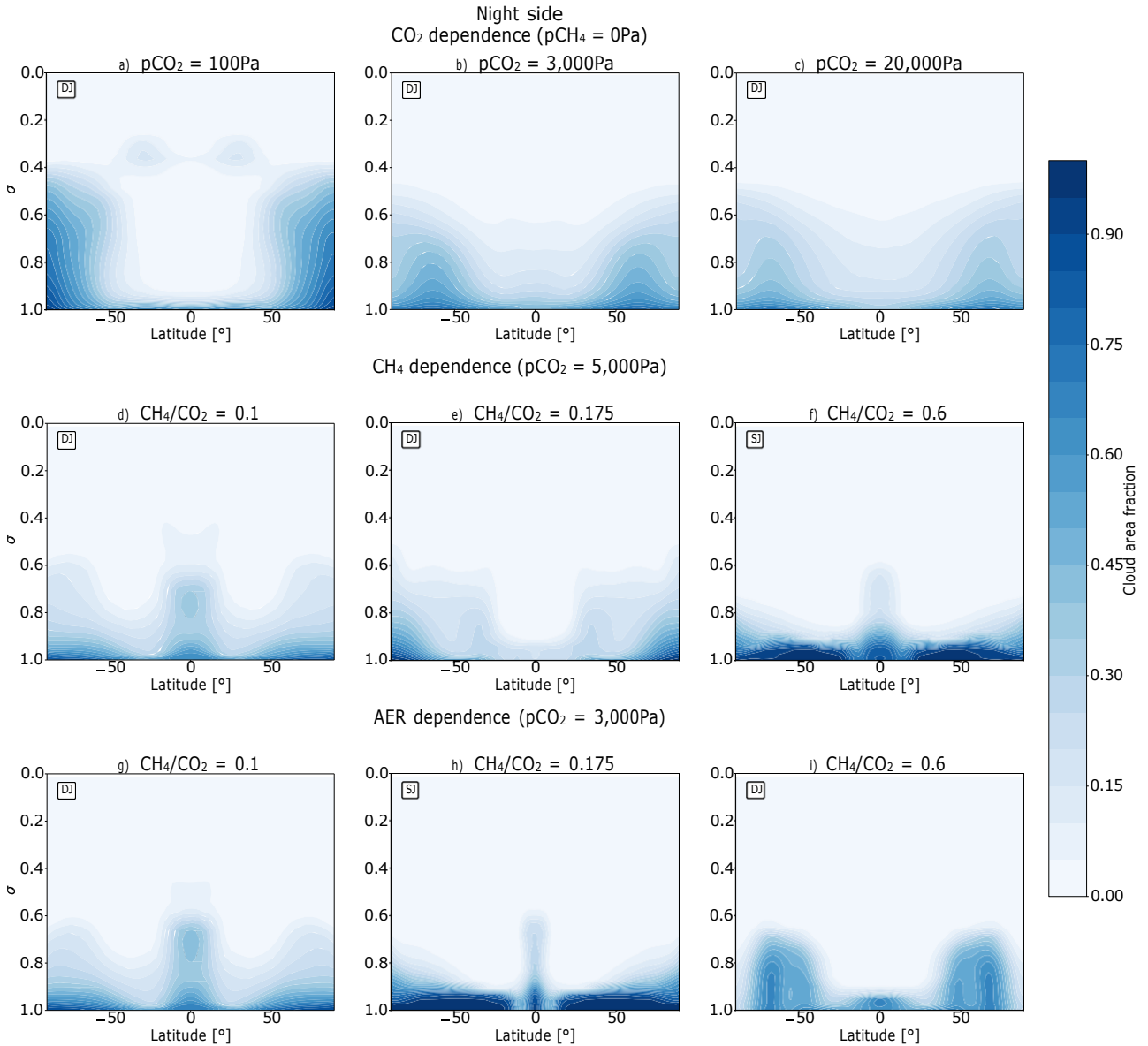

**Figure 13.** Nightside cloud area fraction of the simulations. a, b and c show the noAER simulations with $pCO_2$ = 100, 3,000 and 20,000 Pa ($pCO_2$ fixed at 0 Pa). d, e and f show the noAER simulations with $CH_4/CO_2$ = 0.1, 0.175 and 0.6 ($pCO_2$ fixed at 5,000 Pa). g, h and i show the AER simulations with $CH_4/CO_2$ = 0.1, 0.175 and 0.6 ($pCO_2$ fixed at 3,000 Pa). $\sigma$ (the y-axis) is the air pressure divided by the pressure at the surface.

Here we also compare with the previous works on varying $CH_4$ and $N_2$ concentration for simulations of TRAPPIST-1e. Our results align with the results in Turbet et al. (2018), who found cooling in the lower atmosphere due to the antigreenhouse cooling effect from $CH_4$. Additionally, their results showed surface warming when the concentration of $CH_4$ is very high. This is due to its tropospheric warming dominating over the stratospheric cooling. Nevertheless, the high $CH_4$ concentration used in their work is beyond the limits of our radiative input data. Therefore we are unable to model the effect of having such a high concentration of $CH_4$. In this work we have also seen that the dayside mean surface temperature reaches above 273.15 K with a high concentration of CO$_2$ and a low concentration of CH$_4$. This is similar to the results from Fauchez et al. (2019) in which they adopted the Archean atmospheric configuration from Charnay et al. (2013) in TRAPPIST-1e.

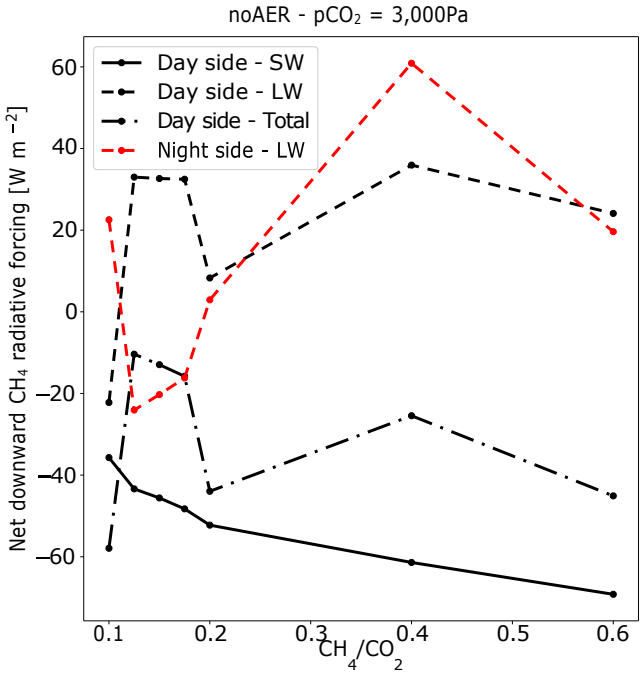

**Figure 14.** Net downward CH$_4$ radiative forcing on the surface in the noAER simulation when pCO$_2$ = 3,000 Pa.

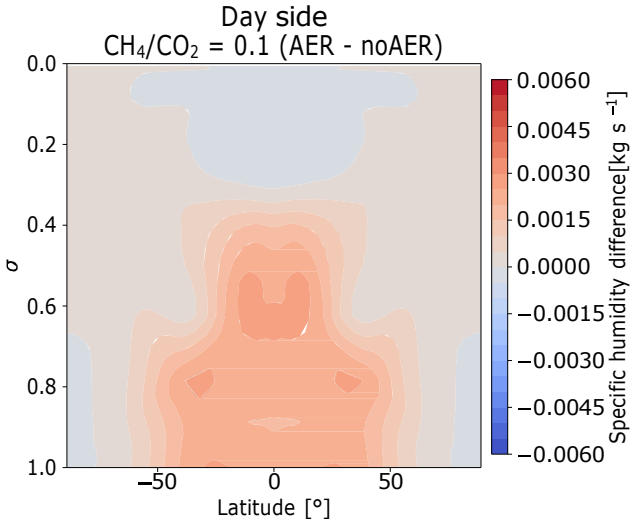

**Figure 15.** Mean zonal difference (AER - noAER) of specific humidity when CH$_4$/CO$_2$ = 0.1 (pCO$_2$ fixed at 3,000 Pa). $\sigma$ (the y-axis) is the air pressure divided by the pressure at the surface.

On the nightside, the surface temperature peaks at a different ratio compared to that found on the dayside for corresponding simulations. For example, in the noAER simulation with pCO$_2$ = 3,000 Pa as discussed above, the surface temperature peaks at pCH$_4$ = 300 Pa (CH$_4$/CO$_2$ = 0.1) on the nightside (Fig 7c). The surface temperature is controlled by the energy transport by the global circulation, more so than by the radiative forcing of CH$_4$. For instance, the CH$_4$ radiative forcing peaks when CH$_4$/CO$_2$ = 0.4, rather than 0.1 (Fig. 14). But when CH$_4$/CO$_2$ = 0.1, the circulation reduces the day-night contrast more efficiently, compared to simulations with other CH$_4$/CO$_2$ ratios. All these effects result on the nightside mean surface temperature is the highest for CH$_4$/CO$_2$ = 0.1. It is difficult to identify a clear trend in the nightside temperature because it depends on the jet regime, which in turn is not linearly dependent on the CO$_2$/CH$_4$ concentration.

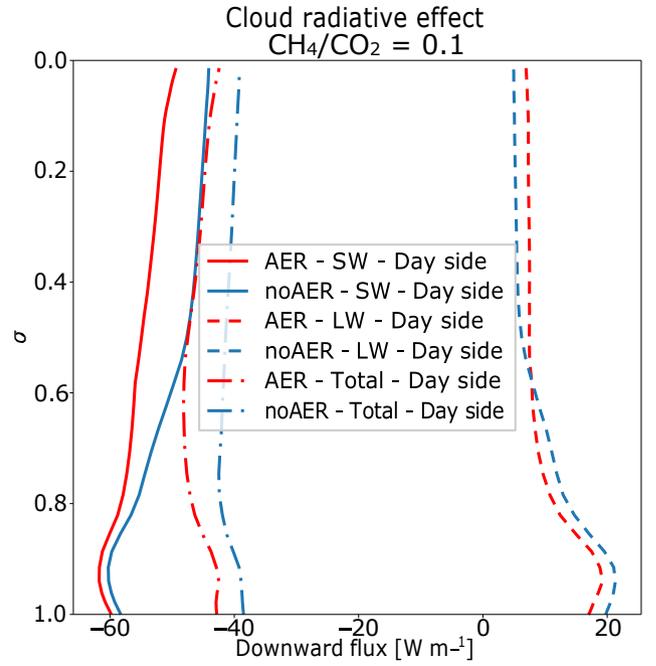

**Figure 16.** Cloud radiative effect of the dayside of both AER and noAER when CH$_4$/CO$_2$ = 0.1 (pCO$_2$ fixed at 3,000 Pa). $\sigma$ (the y-axis) is the air pressure divided by the pressure at the surface.

*4.1.2 Effect of Haze*

In the case of rapid rotation, a thin layer of haze, when CH$_4$/CO$_2$ = 0.1, can lead to global warming of up to ∼10.6 K (Mak et al. 2023). Our work here also shows dayside warming of up to ∼4.9 K (Fig. 7d). Similar to Mak et al. (2023), our work shows that the shortwave heating from the haze warms up the atmosphere, leading to an increase of specific humidity in the atmosphere, as seen in Fig. 15. The increase of water vapour concentration further warms up the atmosphere due to increased absorption of longwave radiation. Nevertheless, contrary to Mak et al. (2023), we found that the presence of haze reduces the cloud distribution in general. This increases radiation loss to space, which can be seen in the radiative cloud feedback plotted in Fig. 16. This change, however, is small, with the combined effect still leading to an increase of the dayside surface temperature.

The nightside of our simulation with CH$_4$/CO$_2$ = 0.1 sees a drop in the surface temperature of ∼10.9 K, when compared to a corresponding haze-free simulation. The increase of the dayside temperature is not replicated on the nightside. This is due to the fact that our thin-haze simulation exhibits a much weaker zonal wind speed of the prograde jets compared to the corresponding noAER simulation (see Fig. 12). When the layer of haze is thick, it acts as a radiation shield and reduces the amount of stellar radiation entering the lower atmosphere. Both the dayside and nightside atmospheres therefore experience a sharp temperature decrease. We thus conclude that the effect of haze on the dayside of a tidally-locked planet is similar to that for a fast rotating planet, regardless of the thickness of the haze layer. The nightside also experiences sharp cooling under the presence of a thick layer of haze as the heat retained in the atmosphere on the dayside has been largely reduced. However, with a thin layer of haze, both night-side cooling and warming are possible depending on the jet structure and its strength.

### 4.2 Future Work

As is the case in Mak et al. (2023), the haze particles are assumed to be spheres in this work and therefore Mie scattering has been applied to calculate the optical properties of the haze particles (Bar-Nun et al. 1988). However, studies have shown that hydrocarbon haze particles are actually fractal agglomerates, with monomers sticking together to form aggregates. These fractal hazes show strong extinction effects in the UV and less in the visible to infrared regime, while spherical hazes show the similar extinction effect in all wavelength (Wolf & Toon 2010). Recent studies involving the inclusion of haze in their 1D and 3D model have also adopted the approach of fractal haze in their simulations (Wolf & Toon 2010; Zerkle et al. 2012; Arney et al. 2016). In particular, Arney et al. (2017) found that fractal hazes result in less cooling around M dwarfs compared to equal mass spherical particles because fractal particles produce less extinction relative to spherical particles at the wavelengths M dwarfs emit most of their energy. This means our results may overestimate both the warming and cooling effect from the haze. In future work, the modified mean-field approximation (Botet et al. 1997; Tazaki & Tanaka 2018; Lodge et al. 2023) will be included in our model.

The photochemical model used here to prescribe the vertical profiles of haze is not coupled to the climate model, and we approximate the haze to be fixed in time and horizontally uniform. However in reality, the circulation would transport haze to different parts of the planet's atmosphere. For example, Wolf & Toon (2010) showed that haze particles are transported from the equatorial region to the polar region in their Early Earth simulations. Kempton et al. (2017) found that in a hot Jupiter atmosphere, haze can be deposited on the nightside due to the superrotating jet and therefore haze is more prone to be observed in the evening terminator. Steinrueck et al. (2021) simulated the transport of haze in the atmosphere of the hot Jupiter HD 189733b. Contrary to Kempton et al. (2017), they showed that small haze particles can be trapped in stationary mid-latitude vortices, resulting in a higher concentration of haze at the morning terminator of the planet. These studies show the importance of a fully interactive haze parameterization, both in terms of the haze production pathways and its transport. This will be the focus of a follow-up work.

## 5 CONCLUSIONS

In this study we simulate TRAPPIST-1e using the 3D GCM, the UM, to understand the impact of varying atmospheric $CH_4$ and $CO_2$, and haze profiles on the climate of a tidally-locked terrestrial planet. We find that changing the atmospheric opacity by varying the concentration of $CH_4$ and $CO_2$, and the thickness of the haze, can lead to either a single jet or double jet regime, with a range of wind speeds. The global circulation then impacts the heat and water vapour and cloud distribution in the atmosphere.

We have found that increasing $pCO_2$ results in a warmer atmosphere on both the day and nightside of the planet due to its greenhouse effect. The increase of surface temperature on the day and nightside reaches up to $\sim 14$ K and $\sim 21.2$ K, respectively. Increasing $pCH_4$ or thickening the haze layer leads to a strong heating in the upper atmosphere on the dayside, due to the shortwave absorption. Alongside this, there is a substantial cooling of the lower atmosphere as there is less shortwave radiation reaching it. However, a thin haze layer, formed when $CH_4/CO_2 = 0.1$, can lead to a dayside warming of $\sim 4.9$ K due to the change in the water vapour distribution. These results are similar to what was found for the rapidly-rotating Archean Earth in Eager-Nash et al. (2023) and Mak et al. (2023). To reach a mean dayside temperature above 273.15 K, our simulations show that the planet's atmosphere should have a higher concentration of $CO_2$, lower concentration of $CH_4$ and a thin layer of haze. The temperature and cloud structure on the nightside of the planet however depends on the circulation regime.

### ACKNOWLEDGEMENTS

We acknowledge funding from the Bell Burnell Graduate Scholarship Fund (grant number BB005), administered and managed by the Institute of Physics, which made this work possible. This work was supported by a UKRI Future Leaders Fellowship [grant number MR/T040866/1], a Science and Technology Facilities Council Consolidated Grant [ST/R000395/1] and the Leverhulme Trust through a research project grant [RPG-2020-82]. Material produced using Met Office Software. We acknowledge use of the Monsoon2 system, a collaborative facility supplied under the Joint Weather and Climate Research Programme, a strategic partnership between the Met Office and the Natural Environment Research Council. This work used the DiRAC Complexity system, operated by the University of Leicester IT Services, which forms part of the STFC DiRAC HPC Facility (www.dirac.ac.uk). This equipment is funded by BIS National E-Infrastructure capital grant ST/K000373/1 and STFC DiRAC Operations grant ST/K0003259/1. DiRAC is part of the National e-Infrastructure. G. Arney acknowledges funding from the Virtual Planetary Laboratory Team, a member of the NASA Nexus for Exoplanet System Science, funded via NASA Astrobiology Program Grant No. 80NSSC18K0829. JE-N would like to thank the Hill Family Scholarship. The Hill Family Scholarship has been generously supported by University of Exeter alumnus, and president of the University's US Foundation Graham Hill (Economic & Political Development, 1992) and other donors to the US Foundation.

### DATA AVAILABILITY

The research data supporting this publication are openly available with CC BY 4.0 Mak et al. (2024).

This paper has been typeset from a T$_E$X/L$^A$T$_E$X file prepared by the author.